\documentstyle[aps,epsf,multicol]{revtex}

\newcommand{\lsim}{\raise.35ex\hbox{$<$}\kern-0.75em\lower.5ex\hbox{$\sim$}}
\newcommand{\gsim}{\raise.35ex\hbox{$>$}\kern-0.75em\lower.5ex\hbox{$\sim$}}
\begin{document}
\draft
\preprint{HEP/123-qed}
\title{Magnetization Process in the One-Dimensional Doped Kondo Lattice Model}
\author{Shinji Watanabe\cite{byline}}
\address{
Department of Physics, Tohoku University, Sendai 980-8587
}

\date{July 28, 1999}

\maketitle

\begin{abstract}
The magnetization process in the one-dimensional Kondo lattice model
for the doped $(n_{c}<1)$ case
is studied by the density matrix renormalization group method.
A rapid increase of the magnetization is caused
by the collapse of the intersite incommensurate correlation of $f$ spins.
On the contrary, the intrasite $f$-$c$ singlet correlation survives
in the larger magnetic field.
The crossover from large to small Fermi surfaces
for majority and minority spins is observed,
whereas the Fermi surfaces are always contributed by $f$ spins.
A magnetization plateau appears with the magnitude of $1-n_{c}$.
Both ends of the plateau are related to the coherence temperature and
the Kondo temperature which are characteristic energies essential
in heavy electron systems.
\end{abstract}


\begin{multicols}{2}

\narrowtext

In heavy electron systems, metamagnetism has been attracting great
interest for a long time.
A typical compound $\rm Ce Ru_{2} Si_{2}$ shows
metamagnetic behavior at the magnetic field
$H_{\rm M}=7.7$ T~\cite{Mignot}.
In a de Haas-van Alphen (dHvA) experiment,
the size of the Fermi surface changes at $H_{\rm M}$~\cite{Aoki}.
According to band structure calculations~\cite{YMGM},
the measured Fermi surface is consistent with
the itinerant $f$-electron band with the large Fermi surface below
$H_{\rm M}$. On the contrary, above $H_{\rm M}$
the measured Fermi surface is small, which is explained by
the band structure of $\rm LaRu_{2}Si_{2}$; $f$ electrons
seem to be localized.
In a neutron measurement, the incommensurate spin correlation
disappears rapidly around $H_{\rm M}$, but spin fluctuation on the single site
survives above $H_{\rm M}$~\cite{RMignot}.

In the course of understanding above physics,
it is a necessary step to exhibit a fundamental mechanism of the
magnetization process in the basic model describing
heavy electron systems.
So far, numerical calculations for small clusters
with a magnetization
have been done in the periodic Anderson model (PAM)
and the Kondo lattice model (KLM)~\cite{Ueda,YamUeda,saso,CuYu,Tutui}.
However, the unified picture of the magnetization process
in the whole range of the magnetic field
has not been established yet even in the basic models.

The purpose of this Letter is to clarify the fundamental
mechanism of the magnetization process in the basic model
on the basis of an accurate method:
We calculate the magnetization, Fermi surfaces and correlation functions
in the one-dimensional KLM with use of
the density matrix renormalization group (DMRG) method~\cite{white1}.
We show that the magnetization plateau is related to two characteristic
energies essential in heavy electron systems.
We also present
the real-space picture and the band picture with respect to
the quasi particle by comparing the magnetization process
of the KLM with that of the PAM.

The one-dimensional KLM in a magnetic field is
\begin{eqnarray}
\label{eq:KLM}
       H =&-&t\sum_{i\sigma}
  (c^{\dagger}_{i\sigma}c_{i+1\sigma}+{\rm H.c.})
 + J\sum_{i}{\bf S}^{f}_{i} \cdot {\bf S}^{c}_{i}
\nonumber \\
&-&h\sum_{i}\left(S^{{f}z}_{i}+S^{{c}z}_{i}\right),
\end{eqnarray}
where $c^{\dagger}_{i\sigma} (c_{i\sigma})$ is a creation
(annihilation) operator
of a conduction electron on the $i$-th site $(1\le i \le L)$
with spin $\sigma$.
${\bf S}^{f}_{i}$ and ${\bf S}^{c}_{i}$ are spin operators
on the $i$-th site of
the $f$ spin and the conduction electron, respectively.
The rescaled magnetic field denoted by $h$
includes the $g$-factor and the Bohr magneton $\mu_{\rm B}$.
The magnetization is defined by
$
m \equiv n^{f}_{\uparrow}-n^{f}_{\downarrow}
        +n^{c}_{\uparrow}-n^{c}_{\downarrow},
$
where $n^{f}_{\sigma}$ and $n^{c}_{\sigma}$ are the densities
of $f$ spins and conduction electrons with spin $\sigma$,
respectively.
The magnetization is related to
$S^{z}_{\rm tot}=\sum_{i}(S^{fz}_{i}+S^{cz}_{i})$ as
$m=2S^{z}_{\rm tot}/L$ and $m$ takes the range of $0 \le m \le 1+n_{c}$.

We study the magnetization process of
eq.~(\ref{eq:KLM}) away from half filling
$(n_{c}=n^{c}_{\uparrow}+n^{c}_{\downarrow}<1)$.
In the DMRG calculation, we typically set the system size $L=40$
under the open boundary condition
and the number of states kept is taken up to 300.

\begin{figure}
\begin{center}
\epsfxsize=8.7cm \epsfbox{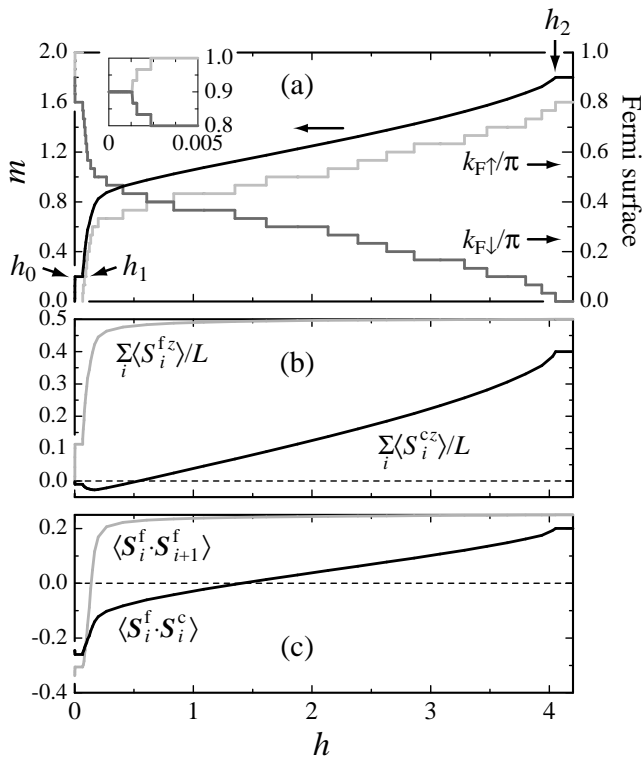}
\end{center}
\caption{(a) Magnetization curve and Fermi surfaces
(b) $\sum_{i}\langle S^{{f}z}_{i} \rangle/L$ and
    $\sum_{i}\langle S^{{c}z}_{i} \rangle/L$
(c) intrasite correlation
     $\langle {\bf S}^{f}_{i} \cdot {\bf S}^{c}_{i}\rangle$
and  intersite correlation
$\langle {\bf S}^{f}_{i} \cdot {\bf S}^{f}_{i+1}\rangle$,
under the magnetic field $h$
for $t=1$ and $J=1$
at $n_{c}=4/5$. The inset in (a) is the enlargement of the small-$h$
regime for $k_{{\rm F}\uparrow}/\pi$ and $k_{{\rm F}\downarrow}/\pi$
(see text).}
\label{fig:nc45m}
\end{figure}

\begin{figure}
\epsfxsize=7.8cm \epsfbox{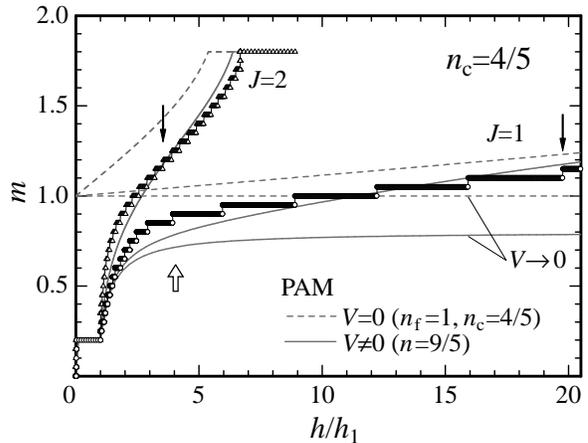}
\caption{
Magnetization of the KLM for $n_{c}=4/5$ and of the PAM for $n=9/5$,
with $t=1$.
The dashed line represents the $V=0$ case of the PAM with $n_{f}=1$
and $n_{c}=4/5$.
Each data for $V \ne 0$ and $V=0$ of the PAM is scaled by $h_{1}^{*}$,
where $h_{1}^{*}$ is set to be equal to $h_{1}$ of the KLM.
The solid arrows show the point where
$\langle {\bf S}^{f}_{i}\cdot{\bf S}^{c}_{i}\rangle$
of the KLM changes the sign.
The open arrow indicates the crossover point
from large to small Fermi surfaces (see text).}
\label{fig:PAM}
\end{figure}

As for the doped case at $h=0$, a ferromagnetic metallic phase is known
to exist in the large-$J/t$ regime~\cite{tune}.
In this Letter
we restrict to the paramagnetic metallic ground state without $h$.
Figure~\ref{fig:nc45m} shows the magnetization process for
$t=1$ and $J=1$ at $n_{c}=4/5$.
A plateau appears with the magnitude of $m=1/5$ as seen near
the left end of Fig.~\ref{fig:nc45m}(a)
(the enlargement will be given in Fig.~\ref{fig:PAM}).
The magnetic field of the left (right) end
of the plateau is denoted by $h_{0} (h_{1})$.
As $n_{\rm c}$ approaches 1, $h_{1}$ increases and continuously connects
to the spin gap $\Delta_{\rm s}$ at half filling without $h$;
$\Delta_{\rm s}$ is associated with the characteristic energy scale called
Kondo temperature $T_{\rm K}$~\cite{tunetugu}.
In order to illustrate the constituents of the magnetization,
the average for the $z$-component of
spin is shown in Fig.~\ref{fig:nc45m}(b).
It turns out that
the main component of the magnetization for $0 \le h \le h_{0}$
comes from $f$ spins.
The rapid increase of $m$ beyond $h_{1}$ is
still dominated by $f$ spins and conduction electrons give the negative
magnetization;
this is due to the internal field via the Kondo-coupling $J$
and note that the density of states shifts to the direction
opposite to that expected in this weak-$h$ regime.
On the contrary, the property for the higher magnetic field  up to
the saturation $h_{2}$ is characterized by conduction electrons.
In Fig.~\ref{fig:nc45m}(c), the intra and intersite spin-spin correlations
for the central $i$-th site are shown.
At $h=0$, the magnetic moments are mostly canceled by
the incommensurate array of $f$ spins, and the residual moments
are screened by formation of the collective Kondo singlet in the
small-$J/t$ regime~\cite{YamUeda,tune}.
Even if $h$ is switched on, the intra and intersite correlations
shows minor change as long as $0 \le h \le h_{1}$.
However, once $h$ exceeds $h_{1}$,
the intersite incommensurate correlation
of $f$ spins is destroyed abruptly to be the ferromagnetic one.
On the contrary, the intrasite $f$-$c$ singlet correlation
survives even in the larger magnetic field.
Consequently, it turns out that the rapid increase of $m$
beyond $h_{1} \sim T_{\rm K}$
is caused by the collapse of the intersite incommensurate correlation
of $f$ spins.

%
%

By the calculations for various $n_{\rm c}$, we find that
the magnetization plateau appears at $m=1-n_{\rm c}$.
This can be intuitively understood if we take the real-space picture
which is relevant in the large-$J/t$ limit:
The number $L(1-n_{c})$ is equivalent to the number of the hole
whose site has an only $f$ spin.
For $0 \le h \le h_{0}$, these $f$ spins
are polarized by the magnetic field, in order to avoid the energy loss
to destroy the intrasite $f$-$c$ singlet correlation.
As long as $h_{0} \le h \le h_{1}$, this state is still stable
so that the magnetization plateau appears.
Once $h$ exceeds $h_{1}$, the $f$-$c$ singlet starts to be broken
as shown above.

We now turn to the location of the Fermi surface.
The Fermi wave number $k_{{\rm F}\sigma}$ for each spin
defined as the singular point
in the momentum distribution function
is detected by the period of the Friedel oscillation~\cite{shibata1}.
The open boundary condition used in the present calculation
works as the perturbation which induces the density oscillations:
The one-particle charge density
for conduction electrons under the open boundary condition
is written as
%
\begin{eqnarray}
\langle N^{c}(x) \rangle
&\sim& A_{\uparrow}
\frac{\cos ( 2 k_{{\rm F}\uparrow} x )}{x^{\gamma_{\uparrow}}}
+A_{\downarrow}
\frac{\cos ( 2 k_{{\rm F}\downarrow} x )}{x^{\gamma_{\downarrow}}}
\nonumber \\
&+&B
\frac{\cos [ 2 (k_{{\rm F}\uparrow}+k_{{\rm F}\downarrow})x]}{x^{\gamma}},
\label{cos2kf}
\end{eqnarray}
%
where $\langle N^{c}(x) \rangle
=\langle N^{c}_{\uparrow}(x) \rangle + \langle N^{c}_{\downarrow}(x) \rangle$
with
$\langle N^{c}_{\sigma}(x) \rangle$ being the one-particle density with
spin $\sigma$~\cite{cardy,Bed}.
The spin density $\langle S^{cz}(x) \rangle
=(\langle N^{c}_{\uparrow}(x) \rangle - \langle N^{c}_{\downarrow}(x)
\rangle)/2$
has the same form as
$\langle N^{\rm c}(x) \rangle$, but with different coefficients
$A_{\sigma}$ and $B$.

%
\begin{figure}
\begin{center}
\epsfxsize=8cm \epsfbox{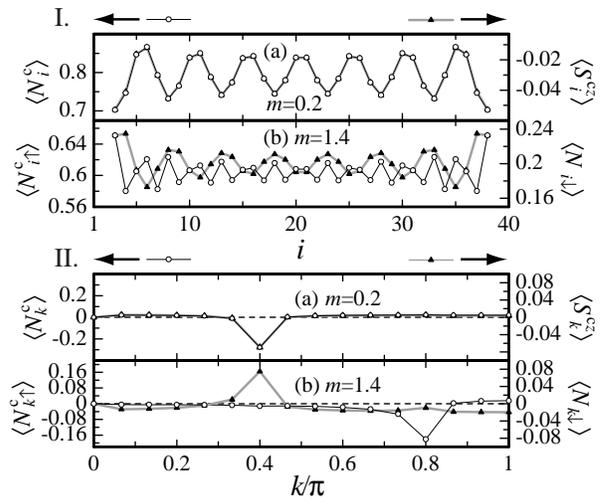}
\end{center}
\caption{
One particle density in the real space (I) and its Fourier spectrum (II).
(a) $m=0.2$ and (b) $m=1.4$ for $t=1$ and $J=2$ at $n_{c}=4/5$.
The Fourier transformation is carried out by using the central 30 sites.
}
\label{fig:nc45ud}
\end{figure}
%

As shown in Fig.~\ref{fig:nc45ud}I,
we calculate
$\langle N^{c}_{i} \rangle =\langle N^{c}_{i\uparrow}\rangle +
\langle N^{c}_{i\downarrow}\rangle$ and
$\langle S^{cz}_{i} \rangle = (\langle N^{c}_{i\uparrow}\rangle-
\langle N^{c}_{i\downarrow}\rangle)/2$
with $\langle N^{c}_{i\sigma}\rangle=
\langle c^{\dagger}_{i\sigma}c_{i\sigma}\rangle$
for each magnetization.
The Fermi wave number is derived from the peak structure in
the Fourier spectrum of
$\langle N^{c}_{i} \rangle$ and $\langle S^{cz}_{i} \rangle$
as in Fig.~\ref{fig:nc45ud}II.
In Fig.~\ref{fig:nc45ud}I(a)
we see that the holes are located mutually keeping the interval of 4 sites
in the case of $m=1/5$ which corresponds to the plateau.
This configuration is detected by the peak structure at
$2(k_{{\rm F}\uparrow}+k_{{\rm F}\downarrow})=2\pi/5$ (mod $2\pi$) in
Fig.~\ref{fig:nc45ud}II(a), which persists in its position even
if $m$ changes.
Additionally, there appears the $2k_{{\rm F}\sigma}$ peak
for each spin whose position depends on $m$.
As $m$ increases $(\gsim 1)$, the peak is detected more easily by
$\langle N^{c}_{i\sigma} \rangle$ rather than by
$\langle N^{c}_{i} \rangle$ and $\langle S^{cz}_{i} \rangle$.
We see that the clear two peaks of
$2k_{{\rm F}\uparrow}$ and $2k_{{\rm F}\downarrow}$ appear and also that the
amplitude for $2(k_{{\rm F}\uparrow}+k_{{\rm F}\downarrow})$ becomes
quite small in Fig.~\ref{fig:nc45ud}II(b).
We show the $J=2$ data in Figs.~\ref{fig:nc45ud}(a) and~\ref{fig:nc45ud}(b),
since the $J=1$ data contain the additional small peak
which is not intrinsic as reported at $h=0$~\cite{shibata1}.
The resultant $k_{{\rm F}\sigma}$ is shown in Fig.~\ref{fig:nc45m}(a).
At $h=0$ the Fermi wave number is given by
$k_{{\rm F}\uparrow}=k_{{\rm F}\downarrow}=\pi(1+n_{c})/2=
9\pi/10$~\cite{shibata1}:
Namely, the Fermi surface is large, which
includes the contribution from $f$ spins.
As $h$ increases from  $0$ to $h_{0}$, $k_{{\rm F}\uparrow}$
$(k_{{\rm F}\downarrow})$ increases (decreases)
to $\pi$ $(\pi n_{c}=4\pi/5)$,
and they remain the same as long as $h_{0} \le h \le h_{1}$.
When $h$ exceeds $h_{1}$, $k_{{\rm F}\uparrow}$ increases from 0  and
$k_{{\rm F}\downarrow}$ decreases from $\pi n_{c}$.
When $h$ reaches the saturation field $h_{2}$,
we obtain $k_{{\rm F}\uparrow}=\pi n_{c}=4\pi/5$ and
$k_{{\rm F}\downarrow}=0$, which are equal to the small Fermi surfaces of the
completely polarized conduction bands decoupled to $f$ spins.

We confirmed that
the Fermi surfaces obtained with use of eq.~(\ref{cos2kf})
and the magnetization derived from the minimization of the total energy
satisfy the following relation;
$
k_{{\rm F}\sigma}/{\pi}=(1+n_{c}\pm m)/2,
$
where
$\sigma=\uparrow (\downarrow)$ corresponds to $+ (-)$ in the right hand side.
This means that there is no phase transition and that
the Luttinger's sum rule holds in a magnetic field.
The expression of $k_{{\rm F}\sigma}$ is consistent with
the argument on the basis of the generalized
Lieb-Schultz-Mattis theorem~\cite{YamOshi}.
The important point is that
the Fermi surface is always {\it large}, which involves the
contribution from $f$ spins, whereas the crossover
from large to small Fermi surfaces is observed~\cite{smallkF}.


In order to analyze the crossover from large to small
Fermi surfaces in detail,
let us consider the PAM:
\begin{eqnarray}
\label{eq:PAM}
H_{\rm PAM}&=&-t\sum_{i\sigma}
\left(
c^{\dagger}_{i\sigma}c_{i+1\sigma}+ {\rm H.c.}
\right)
+\varepsilon_{f}\sum_{i\sigma}n^{f}_{i\sigma}
\nonumber \\
&+&V\sum_{i\sigma} \left( f^{\dagger}_{i\sigma}c_{i\sigma}+
{\rm H.c.} \right)
+U\sum_{i}n^{f}_{i\uparrow}n^{f}_{i\downarrow}.
\end{eqnarray}
The KLM is the effective model derived from the PAM in the regime of
$U \gg \pi \rho(0)V^{2}$
under the symmetric condition $\varepsilon_{f}+U/2=0$,
where $\rho(0)$ is the density of states of conduction electrons
at the Fermi energy.
Finite $U$ makes the effective $f$ level shifted up to the Fermi energy
where the Kondo resonance responsible for
the heavy quasi-particle band arises.
From the viewpoint of the adiabatic continuation, we consider
the simplest case of
$\varepsilon_{f}=-U/2=0$ in eq.~(\ref{eq:PAM}).
In Fig.~\ref{fig:PAM}, the solid curve represents the magnetization of the PAM
with the Zeeman term $-h\sum_{i}(S^{{f}z}_{i}+S^{{c}z}_{i})$
for the doped $(n=n_{f}+n_{c}=9/5)$ case.

For $h_{0}^{*} \le h \le h_{1}^{*}$ the plateau at $m=2-n$
appears as a consequence of the hybridization gap:
$\Delta=-2t+2\sqrt{t^{2}+V^{2}}$.
Here the critical fields $h_{i}^{*}$ $(i=0,1)$ are given by
%
\begin{eqnarray}
h_{i}^{*}&=& t\left\{(-1)^{i}-\cos(n\pi)-(-1)^{i}
\sqrt{1+(V/t)^2} \right.
\nonumber \\
& &+\left. \sqrt{\cos^{2}(n\pi)+(V/t)^{2}}\right\},
\nonumber
\end{eqnarray}
%
and the saturation field $h_{2}^{*}$ by $h_{1}^{*}+t[1+\cos(n\pi)]$.
Namely, the plateau appears as long as the bottom of the upper empty band
with majority spin becomes lower than
the top of the lower filled band with minority spin as $h$ increases.
In order to separate the contribution from conduction electrons,
we also show the magnetization for $V=0$,
setting $n_{f}=1$ and $n_{c}=4/5$.
Additionally,
the discrete data of $J=1$ and 2 in the KLM at $n_{c}=4/5$ are shown,
whose horizontal axis is scaled by $h_{1}$.
To facilitate the comparison between the PAM and the KLM,
the magnetization curve of the PAM is also scaled by $h_{1}^{*}$ and
we choose the hybridization $V$ fixing $t=1$ in eq.~(\ref{eq:PAM}) so that
$h_{1}^{*}$ is equal to $h_{1}$ of the KLM.
The $V=0$ case is scaled by the same unit of the corresponding $V \ne 0$ case.
Note that the data for the PAM are just located on those for the KLM
for $0 \le h \le h_{1}$.
In Fig.~\ref{fig:PAM}, the solid arrow indicates the magnetic field
on which the sign of $\langle {\bf S}^{f}_{i} \cdot {\bf S}^{c}_{i} \rangle$
changes in the KLM (see Fig.\ref{fig:nc45m}(c)).

%
%

As for the PAM, we see  in Fig.~\ref{fig:PAM}
that the magnetization for finite $V$
approaches that for $V=0$ with $n_{f}=1$ and $n_{c}=4/5$ as $h$ increases.
Namely, the crossover from $f$ to conduction electrons which gives the dominant
contribution to the magnetization as well as the Fermi surfaces occurs
beyond $h_{1}^{*}$.
Note that, even in the large-$h$ regime where the Fermi surface is
dominated by conduction electrons,
$f$ electrons are coupled to conduction electrons;
$f$ electrons are always itinerant, but $\it not$ localized.
Since $\Delta$ is reduced to be the order of
$T_{\rm K}=2t \exp\{-U/(8\rho(0)V^{2})\}$ by finite $U$,
the range where $f$ electrons give the main contribution to
the magnetization process
will be drastically reduced  by the many-body effects.
As seen in the $V \to 0$ limit,
the Fermi surface can be close to that of conduction electrons
beyond slightly larger $h$ than $h_{1}^{*} \sim T_{\rm K}$
as indicated by the open arrow.
The difference of the size of the Fermi surface
between $V \to 0$ and $V=0$ cases around the open arrow
becomes small as $n$ approaches 2 (half filling).
Similarly, we expect that the crossover will be observed
around $h_{1}\sim T_{\rm K}$ in the KLM in the realistic $J/t \to 0$
case near half filling, where
the intrasite $f$-$c$ singlet correlation remains.

As seen in Fig.~\ref{fig:PAM},
the general behavior of the PAM is similar to that of the KLM.
This implies that the $k$-space (band)
picture for the PAM continuously connects with
the real-space picture for the KLM explained in Fig.~\ref{fig:nc45m}.
Namely, for the PAM the smaller end of the plateau $(h_{0}^{*})$
is associated with the slope (i.e., velocity) of the lower hybridized band
whose component is dominated by $f$ electrons, and
the larger one $(h_{1}^{*})$ connects to the hybridization gap $\Delta$
as $n$ approaches to 2.
The correspondence between the above picture and the result of the KLM
leads us to find the following meaning of the magnetization plateau in
the KLM:
The smaller end of the plateau $(h_{0})$ is associated with the velocity
of the effective carriers whose main component is unscreened $f$ spins
with density $1-n_{c}$, and
the larger one $(h_{1})$ is related to the intrasite $f$-$c$ singlet
correlation, which connects to the spin gap $\Delta_{\rm s}$
as $n_{c}$ approaches 1.
These two characteristic energies are considered to be related to the
coherence temperature $T^{*}$ and the Kondo temperature
$T_{\rm K}$; the former and the latter are often observed as the
characteristic temperatures which give $T^{2}$ and ${\log}T$ behaviors
in the resistivity, respectively in heavy electron compounds.
The quasi-particle picture with respect to the KLM obtained here via
the response to the magnetic field is consistent with recent calculations
of dynamical and finite-temperature quantities without $h$
by N. Shibata, {\it et~al}.~\cite{shibataT}

When realistic factors which should enter are taken into account,
the plateau tends to be smeared;
the difference of $g$ factors between $f$ and conduction electrons,
anisotropy of the hybridization, the dispersion of the
$f$ band, etc.~\cite{Evans}
Hence, the shape of $m$ itself for $0\le h \le h_{1}$ changes
according to these factors.
However, physics shown here offers significant pictures
to understand the magnetization process of heavy electron compounds.
As for $\rm CeRu_{2}Si_{2}$, it was confirmed that there is no
first order phase transition at $H_{\rm M}$ and the electronic state
changes continuously~\cite{sakakibara}.
This suggests that the crossover from large to small Fermi surfaces
occurs with the $f$-electron number involved.
Simultaneously, we pointed out that the localization
of $f$ electrons above $H_{\rm M}$ is not necessary to interpret
both the results of the dHvA experiment~\cite{Aoki}
and the band structure calculations~\cite{YMGM}.
Furthermore, we show the rapid increase of the magnetization due to
the collapse of the intersite correlation of $f$ spins and the
toughness of the intrasite $f$-$c$ singlet correlation.
These tendencies are
observed by the neutron measurement~\cite{RMignot}.

In conclusion, we show the fundamental mechanism of the magnetization
process in the one-dimensional KLM
and shed light on the heavy quasi-particle picture
via the response to the magnetic field.


The author would like to thank Y. Kuramoto for bringing my attention
to this problem as well as the fruitful discussions.
The author is grateful to N. Shibata for sending his data prior
to the publication as well as useful discussions.
Thanks are also due to H. Tsunetsugu, H. Yamagami,
M. Oshikawa and O. Sakai for valuable discussions.
A part of the numerical calculation was performed by VPP500 at the
Supercomputer Center of the ISSP, Univ.~of Tokyo.


\end{multicols}


\begin{references}
\bibitem{smallkF} At the saturation with $m=1+n_{c}$, the Fermi surfaces
can be equal to the small ones;
$2k_{{\rm F}\uparrow}=2\pi(1+n_{c})=2\pi n_{c}$ and
$2k_{{\rm F}\downarrow}=0$.
\end{references}
\end{document}